\documentclass[prx,showpaces,preprintnumbers,amsmath,amssymb,twocolumn,superscriptaddress]{revtex4-1}
\usepackage{graphicx}
\usepackage{dcolumn}
\usepackage{bm}
\usepackage{amssymb}
\usepackage{amsmath}
\usepackage{epstopdf}
\usepackage{booktabs}
\usepackage{threeparttable}
\usepackage[pdfstartview=FitH, CJKbookmarks=true, bookmarksnumbered=true, bookmarksopen=true, colorlinks, pdfborder=001, linkcolor=blue, anchorcolor=blue, citecolor=blue]{hyperref}

\begin{document}
\title{Superconductivity and structural distortion in BaPt$_2$As$_2$}
\author{W. B. Jiang}
\affiliation{Center for Correlated Matter and Department of Physics, Zhejiang University, Hangzhou, 310027, China}
\author{C. Y. Guo}
\affiliation{Center for Correlated Matter and Department of Physics, Zhejiang University, Hangzhou, 310027, China}
\author{Z. F. Weng}
\affiliation{Center for Correlated Matter and Department of Physics, Zhejiang University, Hangzhou, 310027, China}
\author{Y. F. Wang}
\affiliation{Center for Correlated Matter and Department of Physics, Zhejiang University, Hangzhou, 310027, China}
\author{Y. H. Chen}
\affiliation{Center for Correlated Matter and Department of Physics, Zhejiang University, Hangzhou, 310027, China}
\author{Y. Chen}
\affiliation{Center for Correlated Matter and Department of Physics, Zhejiang University, Hangzhou, 310027, China}
\author{G. M. Pang}
\affiliation{Center for Correlated Matter and Department of Physics, Zhejiang University, Hangzhou, 310027, China}
\author{T. Shang}
\affiliation{Center for Correlated Matter and Department of Physics, Zhejiang University, Hangzhou, 310027, China}
\author{X. Lu}
\affiliation{Center for Correlated Matter and Department of Physics,
Zhejiang University, Hangzhou, 310027, China}
\affiliation{Collaborative Innovation Center of Advanced
Microstructures, Nanjing University, Nanjing 210093, China}
\author{H. Q. Yuan}
\email{hqyuan@zju.edu.cn}
\affiliation{Center for Correlated Matter
and Department of Physics, Zhejiang University, Hangzhou, 310027,
China}
\affiliation{Collaborative Innovation Center of Advanced
Microstructures, Nanjing University, Nanjing 210093, China}

\begin{abstract}

We report the synthesis of BaPt$_2$As$_2$ single crystals and the discovery of superconductivity and a structural phase transition in this compound by measuring the electrical resistivity, magnetic susceptibility and specific heat as well as the X-ray diffraction at low temperatures. BaPt$_2$As$_2$ crystallizes in the CaBe$_2$Ge$_2$-type tetragonal structure (P4/nmm) at room temperature and undergoes a first-order structural transition at $T_S\simeq 275$ K, which is likely associated with a charge-density-wave (CDW) instability. BCS-like superconductivity with two subsequent transitions $T_{c1}\simeq1.67$ K and $T_{c2}\simeq1.33$ K are observed. Our results demonstrate that BaPt$_2$As$_2$ may serve as a new system for studying the interplay of superconductivity and the CDW order.

\begin{description}
\item[PACS number(s)]
 74.70.-b, 71.45.Lr, 74.10.+v
\end{description}
\end{abstract}
\maketitle

\section{Introduction}

A growing number of experimental evidences have demonstrated that unconventional
superconductivity may emerge on the verges of magnetic, charge
and valence instabilities~\cite{Mathur1998,Chen2008,Morosan2006,yuan2003}.
Understanding the interplays between superconductivity and its
competing density-wave orders is fundamentally important for
revealing the pairing mechanisms of unconventional superconductors,
e.g., the high $T_c$ cuprates, heavy fermions systems and the
recently discovered Fe-based superconductors. Charge density
wave, a collective state coupled with periodic lattice
distortion, develops in numerous systems. In early days, intensive
efforts were devoted to the studies of superconductivity (SC) and
CDW order in some low-dimensional electronic systems, e.g., the layered chalcogenide compounds NbSe$_3$ and 2H-TaSe$_2$~\cite{Gabovich2001}. It was argued that superconductivity competes with the CDW state in these compounds due to the gap opening on the Fermi surface of a CDW instability, but a detailed study of this competing interaction is still lacking. In a more recent work, it
was shown that superconductivity emerges while suppressing the CDW state in Cu$_x$TiSe$_2$~\cite{Morosan2006}, in a way similar to superconductivity near a magnetic instability~\cite{Mathur1998}. Furthermore, recent
experiments provided strong evidences that superconductivity is
closely associated with a CDW state in the high $T_c$ cuprates
~\cite{Chang2012,Darius2013}. To further explore the intricate relationship
between the SC and CDW states, it is desirable to search for
new types of CDW superconductors.

Recently, a class of $5d$ transition-metal pnictides attracted
considerable interests. SrPt$_2$As$_2$ undergoes a structural phase
transition around $T_s\approx 470$K, changing from a
high-temperature tetragonal phase with the CaBe$_2$Ge$_2$-type
structure (space group P4/nmm) to a phase with an average structure
of the orthorhombic space group Pmmn upon decreasing temperature
~\cite{Imre2007}. A superstructure with a modulation vector
$\textbf{q} = 0.62\textbf{a*}$ was observed in the low-temperature
phase, suggesting the formation of a CDW state below $T_s$
~\cite{Imre2007,Fang2012}. Recent measurements of optical
conductivity revealed a possible energy gap associated with the
structural distortion, providing a further evidence to the CDW
transition at $T_s$~\cite{Fang2012}. More interestingly,
superconductivity shows up while cooling down below
$T_c=5.2$ K~\cite{Kudo2010}. Measurements of heat capacity suggest
two-band BCS superconductivity for SrPt$_2$As$_2$~\cite{Xu2013}.
Similarly, superconductivity and a possible CDW state were also observed in the iso-structural sister
compound SrPt$_2$Sb$_2$~\cite{Motoharu2013}. It was argued that
superconductivity may coexist with the CDW state in these $5d$ transition-metal
pnictides~\cite{Kudo2010}, which requires more careful explorations.

In this article, we report the discovery of superconductivity in a possible new CDW compound
BaPt$_2$As$_2$. BaPt$_2$As$_2$ crystallizes in the
CaBe$_2$Ge$_2$-type tetragonal structure at room temperature. Upon
cooling down, the compound undergoes a pronounced first-order
transition at $T_s\simeq 275$ K as indicated in the electrical resistivity
$\rho_s(T)$ and the specific heat $C(T)/T$. Measurements of X-ray
diffraction (XRD) confirm a structural distortion at $T_s\simeq 275$ K,
suggesting a possible CDW transition as previously shown in
SrPt$_2$As$_2$~\cite{Imre2007,Fang2012}. BCS-like superconductivity shows up at $T_c\simeq 1.7$K, well below the structural phase transition. These
findings suggest that BaPt$_2$As$_2$ may serve as a suitable new system for systematically studying the interplay of superconductivity and CDW in the presence of strong spin-orbit coupling.

\section{Experimental methods}

Single crystals of BaPt$_2$As$_2$ were grown by a self-flux method.
The precursor PtAs$_2$ was first fabricated by heating the Pt
powder ($99.9$+\%) and As grains ($99.999\%$) up to 700$^\circ$C; all the
raw materials used in this study were from Alfa Aesar. The obtained
PtAs$_2$ was mixed with Ba ingots ($99$+\%) and Pt powders according
to an equal molar ratio, and then installed in an alumina crucible which
was sealed in an evacuated quartz ampule. The mixture was heated up
slowly to 1100$^\circ$C and kept at this temperature for 12 hours.
Then it was cooled down to 950$^\circ$C at a rate of 1.5$^\circ$C/h,
followed by a furnace cooling. Shiny plate-like crystals, with a
typical dimension of
$1\textrm{mm}\times 1\textrm{mm}\times0.2\textrm{mm}$, were obtained
by mechanically isolating them from the excess flux. The crystal
structure was characterized by X-ray diffraction (XRD) on a PANalytical
X'Pert MRD diffractometer with Cu K$\alpha$ radiation and a
graphite monochromator. The chemical compositions were
determined by using an energy dispersion X-ray spectroscopy(EDX).
The electrical resistivity $\rho(T)$ and heat capacity $C(T)$ were
measured by a commercial Physical Property Measurement System (PPMS-14T)
equipped with a $^3$He insert down to 0.4 K. Magnetic
susceptibility were measured by using a SQUID magnetometer down to
2K (MPMS-5T, Quantum Design) and
an $ac$ magnetometer in a He-3 refrigerator down to 0.3 K, respectively.

\begin{figure}[h]
\centering\includegraphics[width=7.5cm]{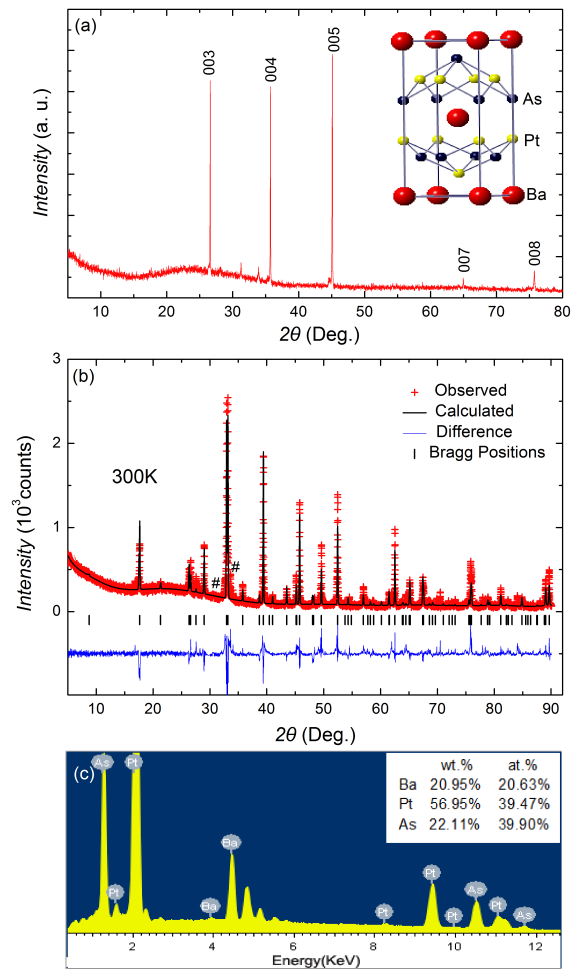}
\caption{(Color online) (a)The XRD patterns of BaPt$_2$As$_2$ single crystals at $T=$300 K. (b) Powder XRD patterns of the grounded BaPt$_2$As$_2$ crystals at $T$=300 K, together with the Rietveld refinement fits (solid black line), the Bragg peaks (vertical bars) and the difference profiles (solid blue line). The unindexed peaks marked by $\#$ are derived from a tiny amount of BaPt$_4$As$_6$ crystals mixed in BaPt$_2$As$_2$ crystals. (c) Energy-dispersion spectroscopy of BaPt$_2$As$_2$ single crystals; the derived atomic and weight ratios of Ba:Pt:As are also shown.}
\label{fig1}
\end{figure}

\begin{figure}[h]
\centering\includegraphics[width=7.5cm]{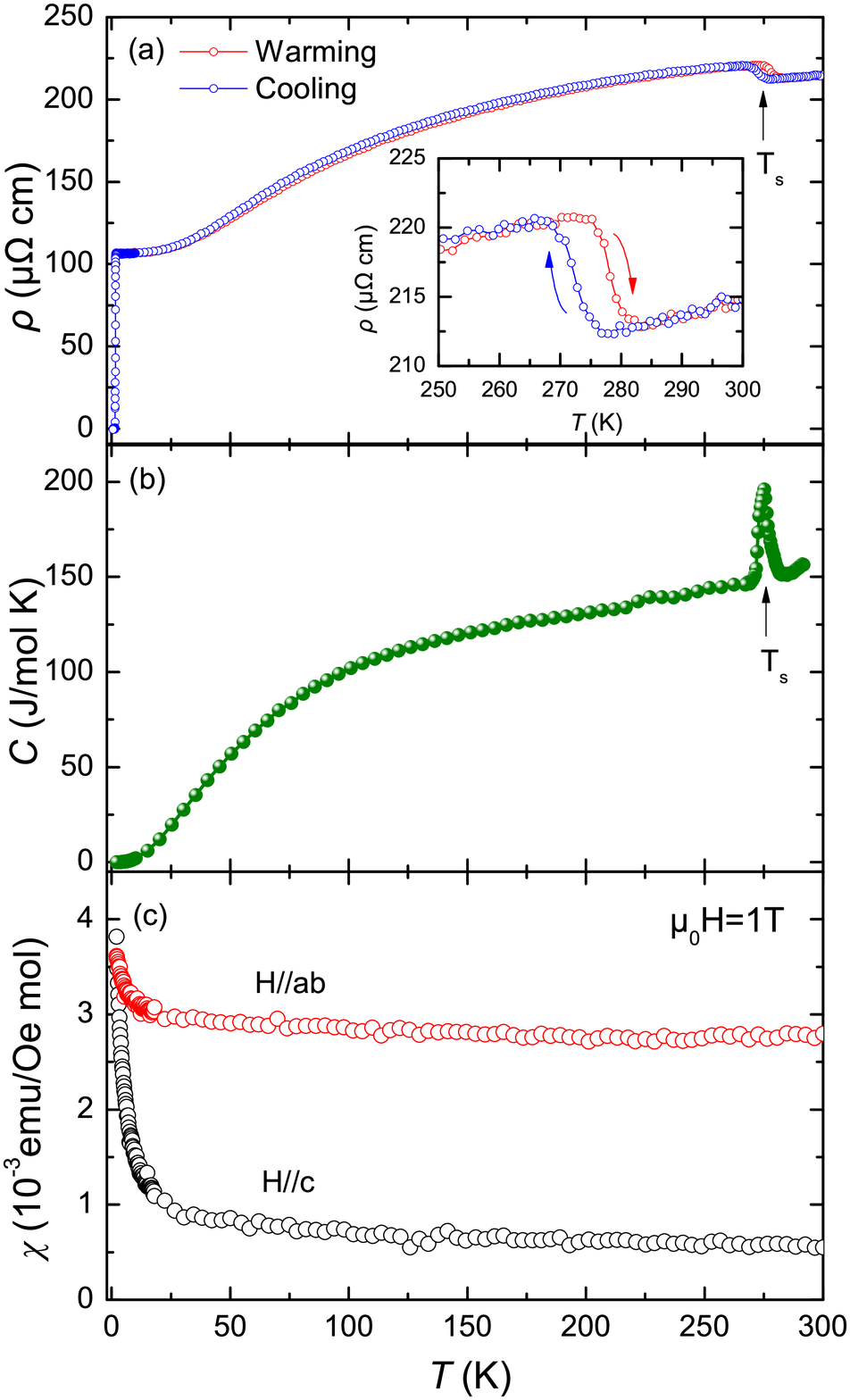}
\caption{(Color online) Temperature dependence of (a) the in-plane electrical resistivity $\rho(T)$ measured
in the warming-up and cooling-down processes, respectively; (b) the heat capacity $C(T)$ and (c) the magnetic susceptibility $\chi(T)$ for BaPt$_2$As$_2$ (sample \#A). A thermal hysteresis of the electrical resistivity at $T_s$ is shown in the inset of (a). A field of 1T was applied for the measurements of $\chi(T)$. }
\label{fig2}
\end{figure}

\section{Results and discussion}

Figure 1(a) shows the XRD patterns of BaPt$_2$As$_2$ single crystals at 300K. The observed (00$l$) ($l$=integer) diffraction peaks suggest that the large crystal planes are perpendicular to the $c$-axis. The narrow full width at half maximum (FWHM) of the
diffraction peaks ($<0.07^\circ$) demonstrates a good sample quality. To further determine the crystal structure, we ground the BaPt$_2$As$_2$ crystals and performed the powder XRD measurements. The obtained XRD patterns, together with the calculated profiles, are shown in Figure 1(b). The experimental diffraction peaks can be well indexed by the CaBe$_2$Ge$_2$-type tetragonal structure with a space group P4/nmm. Note that a few unindexed peaks marked by $\#$ in the figure are likely attributed to a small amount of BaPt$_4$As$_6$ impurity phase mixed in the BaPt$_2$As$_2$ crystals. The Rietveld refinement fits give the lattice parameters of $a=b=4.564{\AA}$ and $c=10.02{\AA}$, in a good agreement with the reference data~\cite{Imre2007}. The derived crystal structure of BaPt$_2$As$_2$ is shown in the inset of Fig. 1(a). The actual sample compositions are also analyzed with the EDX method. We examined a number of crystals from different growths, and consistently obtain an atomic ratio of Ba:Pt:As close to 1:2:2. As an example, we show one of the EDX results in Fig. 1(c); the derived compositions are nearly the same as the stoichiometric BaPt$_2$As$_2$, ensuring the right compound obtained.

Figure 2 presents the physical properties of BaPt$_2$As$_2$ at high temperatures. The electrical resistivity $\rho(T)$, measured with a current applied parallel to the ab-plane, shows a jump around $T_s\simeq$ 272 K upon cooling down from room temperature (see Fig. 2a). A resistivity hysteresis is observed between the cooling-down and warming-up processes, characterizing a first-order transition at $T_s$. The resistive increase at $T_s$ is likely attributed to a gap opening associated with the formation of a CDW order at low temperatures (see below). Such a first-order transition is further demonstrated in the temperature dependent specific heat $C(T)$, where a $\delta$-like anomaly is observed around 275K (see Fig. 2b). On the other hand, no clear transition is seen in the magnetic susceptibility $\chi(T)$ as shown in Fig. 2c, suggesting that the transition at $T_s$ cannot be of a magnetic origin. One possibility is that the changes of charge susceptibility caused by the CDW transition is negligible in comparison with the total susceptibility so that one cannot detect its change at $T_s$. Overall, the magnetic susceptibility $\chi(T)$ shows weak temperature dependence with a certain anisotropy, where the susceptibility for $H\parallel c$, $\chi_c(T)$, is about three times larger than that for $H\parallel ab$, $\chi_{ab}$. The Curie-Weiss-like increase of $\chi(T)$ at low temperatures is likely due to a tiny amount of magnetic impurities.

\begin{figure}[h]
\centering\includegraphics[width=7.5cm]{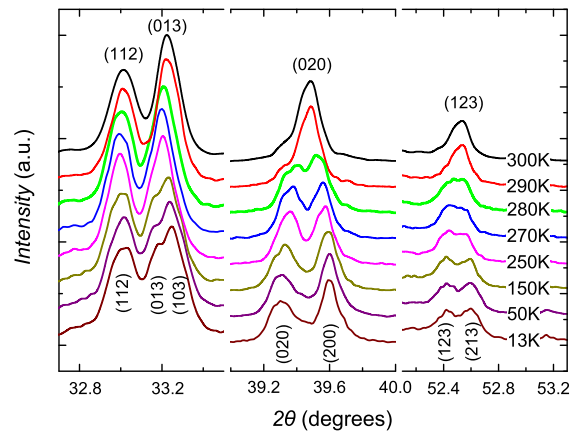}
\caption{(Color online) Powder X-ray diffraction of BaPt$_2$As$_2$ at various temperatures. The diffraction peaks of (013),(020) and (123) are clearly split into two peaks below $T_S\simeq 280$ K.}
\label{fig3}
\end{figure}

\begin{figure}[h]
\centering\includegraphics[width=7.5cm]{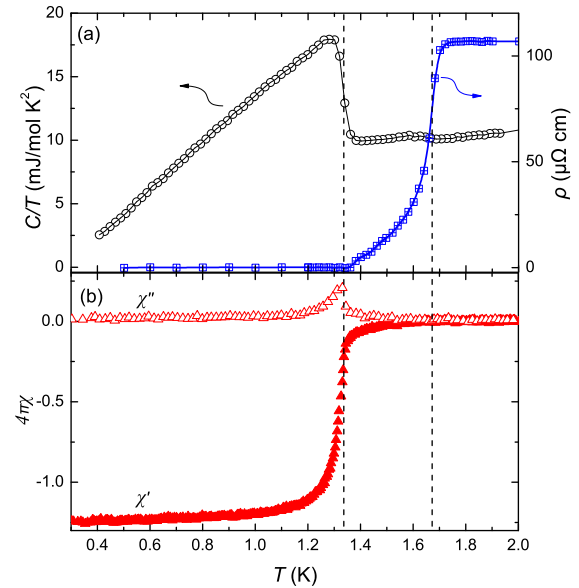}
\caption{(Color online) Temperature dependence of (a) the electrical resistivity $\rho(T)$ and specific heat $C(T)$, and (b) the $ac$ magnetic susceptibility $\chi(T)$ of BaPt$_2$As$_2$ (sample \#B). Here $\chi'$ and $\chi''$ represent the in-phase and out-of phase signals, respectively.
The dashed lines mark the positions of the two superconducting transitions at $T_{c1}=1.67$ K and $T_{c2}=1.33$ K. }
\label{fig4}
\end{figure}

To characterize the transition at $T_s$ in more details, we performed the powder XRD measurements at various temperatures. As described above, the diffraction patters at room temperature can be well indexed by the CaBe$_2$Ge$_2$-type tetragonal structure. With deceasing temperature, a number of diffraction peaks show clear splitting below about 280K. For example, Fig.3 shows a few diffraction peaks which undergo a pronounced change across $T_s$. This fact provides direct evidence that the first-order transition at $T_s\simeq 275$ K, both observed in the electrical resistivity and the specific heat, is caused by a structural change. The resulting diffraction peaks at low temperatures are consistent with the orthorhombic structure (space group Pmmn), the same as the low-temperature phase identified in SrPt$_2$As$_2$~\cite{Imre2007}. This corresponds to a lattice distortion with a shrink along the \textit{a}-axis and an expansion along the \textit{b}-axis from the high-temperature tetragonal phase. Note that the resolution of the powder XRD obtained in the tiny amount of BaPt$_2$As$_2$ crystals at low temperatures is not sufficient to accurately determine the temperature dependence of the lattice parameters. A similar structural phase transition from a tetragonal to an orthorhombic phase upon decreasing temperature was previously observed in SrPt$_2$As$_2$~\cite{Imre2007} and some iron-based superconductors~\cite{Ni2008,Rotter2008,Luo2009}. In iron pnictides, the structural transition is usually followed or accompanied by a spin-density-wave transition~\cite{Rotter2008}. On the other hand, evidence for a structural modulation with $\textbf{q} = 0.62\textbf{a*}$ was shown in SrPt$_2$As$_2$~\cite{Imre2007,Fang2012}, suggesting the formation of a CDW state below $T_s=470$ K. We would thus argue that the structural distortion in BaPt$_2$As$_2$ may correspond to a similar CDW transition, whose nature requires more experimental investigations.

\begin{figure}[ht]
\centering
\includegraphics[width=7.5cm]{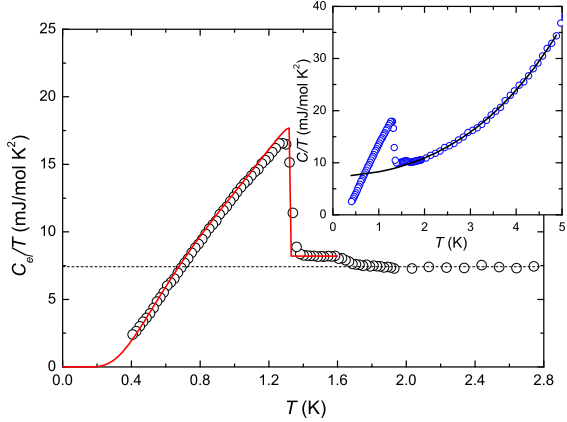}
\caption{(Color online) Temperature dependence of the heat capacity of BaPt$_2$As$_2$ (sample \#B). The main panel plots the electronic specific heat $C_e(T)/T$ versus $T$ after subtracting the lattice contributions. The solid line presents a fit of the conventional BCS model with a gap amplitude of $\Delta=1.73k_BT_c$ at zero temperature. The inset shows the total specific heat $C(T)/T$ and a polynomial fit of $C(T)=\gamma T+B_3T^3+ B_5T^5$ (solid line). }
\label{fig5}
\end{figure}

More interestingly, BaPt$_2$As$_2$ becomes superconducting when lowering temperature further. In Fig. 4, we present the electrical resistivity $\rho(T)$, specific heat $C(T)/T$ and \textit{ac} magnetic susceptibility $\chi_{ac}(T)$ at low temperatures. The resistivity $\rho(T)$ shows a superconducting onset around 1.69 K and reaches zero value around 1.35 K. We note that some samples demonstrate a step-like resistive transition at $T_{c2}$. The broad tail of the superconducting transition is attributed to two subsequent superconducting transitions as seen in the specific heat and magnetic susceptibility data. The specific heat $C/T$ shows a small jump at $T_{c1}\simeq 1.67$ K and another dominant jump at $T_{c2}\simeq 1.33$ K, where the two transition temperatures are close to the onset and zero value of the resistive transition, respectively. Similar features of two superconducting transitions are also seen in the diamagnetic susceptibility $\chi_{ac}(T)$, where the in-phase $\chi^{'}(T)$ and the out-of-phase susceptibility $\chi{''}(T)$ become separated around $T_{c1}$, followed by a sharp decrease of $\chi{'}(T)$ around $T_{c2}$. Below $T_{c2}$, the diamagnetic susceptibility is saturated at a value of $4\pi\chi^{'}\simeq-1.2$. The magnetic shielding volume beyond 100\% is likely attributed to the absence of corrections by the demagnetizing factor. As shown below, the upper critical field $\mu H_{c2}(T)$ of the two superconducting transitions can be well scaled in the same behavior, probably with the same pairing states. We stress that the two superconducting transitions are observed in a number of samples synthesized with different growing conditions, showing that the two subsequent superconducting transitions are well reproducible. It is noted that similar multiple superconducting transitions were previously reported in other compounds, e.g., the heavy-fermion superconductor PrOs4Sb12 ~\cite{Vollmer2003}. Nevertheless, at this stage one also cannot exclude the possibility of filamentary superconductivity at $T_{c1}$ caused by, e.g., sample inhomogeneities, since the changes of magnetic susceptibility and specific heat at $T_{c1}$ are weak and the resistivity drops to zero just below $T_{c2}$.

\begin{figure}[ht]
\centering\includegraphics[width=7.5cm]{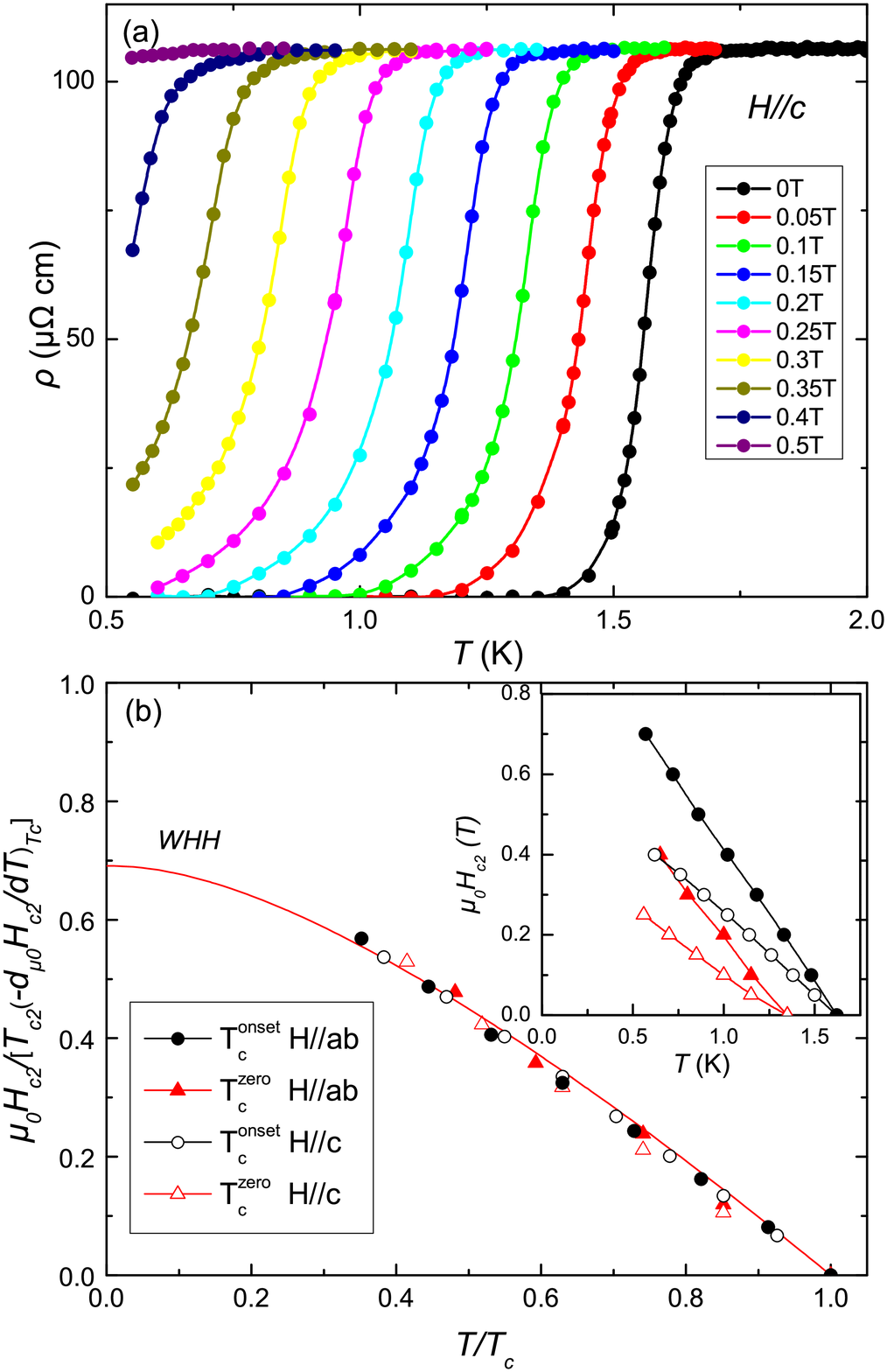}
\caption{(Color online) The upper critical field $\mu_0 H_{c2}$ of BaPt$_2$As$_2$ (sample \#A). (a) Temperature dependence of the electrical resistivity $\rho(T)$ at various magnetic fields ($H//ab$). (b) The normalized upper critical fields versus $T/T_c$ for the two superconducting transitions, which can be nicely fitted by the WHH model (solid line). The inset shows the upper critical fields $\mu_0H_{c2}(T)$, determined at the onset points (circles) and the zero resistance (triangles) of the resistive transition. }
\label{fig6}
\end{figure}

In order to gain insights into the superconducting pairing states, here we analyze the electronic specific heat $C_e(T)/T$ at low temperatures. In the inset of Fig. 5, we fit the specific heat in the normal state by a polynomial expression, i.e., $C=\gamma T+\beta_3 T^3+\beta_5 T^5$, where $\gamma$ is the Sommerfeld coefficient, and the latter two terms represent the lattice contributions. The fits yield $\gamma$=7.42mJ/mol K$^2$, $\beta_3=0.823$ mJ/mol K$^4$ and $\beta_5=0.0132$ mJ/molK$^6$. A Debye temperature of $\Theta_D\simeq$ 227K is estimated by $\theta_D=\sqrt[3]{12\pi^4nR/5\beta_3}$, where $n=5$ is the number of atoms in each unit cell and $R=$8.314J/mol K. The electronic specific heat $C_e(T)/T$, obtained after subtracting the phonon contributions as shown above, is plotted in Fig. 5. Here we will focus on the specific heat jump at $T_{c2}$ since it shows a much weaker jump at $T_{c1}$. The specific heat jump at $T_{c2}$ is determined to be $\Delta C/\gamma T_c$=1.27, being slightly smaller than the BCS value of 1.43. It is noted that inclusion of the anomaly at $T_{c1}$ in the total specific heat jump gives $\Delta C/\gamma T_c\simeq 1.37$ at the low-temperature superconducting transition. A reduced specific heat jump at $T_c$, in comparison with the BCS value, was previously observed in the multiband superconductors, e.g., the well-known MgB$_2 $~\cite{Bouquet2001}. Furthermore, a moderate electron-phonon coupling strength of $\lambda$=0.496 is estimated from the McMillan formula assuming a Coulomb pseudopotential $\mu^*$=0.13 ~\cite{Mcmillan1968}. In the superconducting state, the specific heat $C_e/T$ can be fitted by a weak-coupling BCS model with an energy gap of $\Delta_0=1.73k_BT_c$ at zero temperature (see Fig. 5). In order to precisely determine the gap structure, further measurements down to lower temperatures are needed.

In Fig. 6(a), we present the temperature dependent electrical resistivity $\rho(T)$ at various magnetic fields. Here the field is applied along the $c$ axis and similar results are also obtained for $H//ab$. With increasing magnetic field, the superconducting transition is eventually suppressed to lower temperatures, and the resistive tail becomes more pronounced. As stated above, the onset and zero resistive transitions correspond to two superconducting transitions at $T_{c1}$ and $T_{c2}$, respectively. In the following, we determine the upper critical fields of the two superconducting states by tracking the field dependence of the resistive onset and the zero resistance, which values are shown in the inset of Fig. 6(b). The upper critical field $\mu_0H_{c2}(T_c)$ shows certain anisotropy for the two perpendicular field orientations, and the anisotropic parameter, defined as $\nu=H_{c2}^{ab}/H_{c2}^c$, is nearly temperature independent over the range we measured  ($\nu\simeq1.6$). Such an anisotropic behavior resembles that of some iron-based superconductors~\cite{ZJL2011}. Following the Werthamer-Helfand-Hohenberg (WHH) formula~\cite{WHH1966}, one can estimate the upper critical field at zero temperature from the initial slope of $\mu_0H_{c2}(T)$ near $T_c$, i.e., $\mu_0H_{c2}(0)=0.693T_c\ \textrm{d}\mu_0H_{c2}(T)/\textrm{d}T$$\mid_{T=T_c}$, where the derived values of $\mu_0H_{c2}(0)$ are listed in Table.1. The Ginzburg-Landau coherence length can be estimated by $\xi=(\frac{\phi_0}{2\pi\mu_0H_{c2}})^{1/2}$, which are also summarized in Table.1. In Fig. 6(b), we  plot the normalized upper critical field, $\mu_0H_{c2}/[T_c(d\mu_0H_{c2}/dT)_{T_c}]$, as a function of $T/T_c$ for both $H//c$ and $H//ab$. One can see that the upper critical fields, determined from either the onset point or the zero resistance, follow the same scaling behavior in the two field directions, suggesting that the two superconducting transitions at $T_{c1}$ and $T_{c2}$ originate from the same pairing mechanism.

\begin{table}
  \centering
  \caption{The superconducting parameters for BaPt$_2$As$_2$ which were derived from the resistivity data $\rho(T,H)$ of sample $\#$A.}
  \begin{tabular}{*6c}
  \hline\hline
    & $T_c$ & $-\frac{d\mu_0H_{c2}}{dT}\mid_{T_c}$ & $\mu_0H_{c2}(0)$ & $\xi_{GL}(0)$ & Field direction \\
    & (K) & ($\textrm{T}/\textrm{K}$) & (T) & $({\AA})$ &  \\ [0.5ex]
   \hline
    &  & 0.76 & 0.85 & 197 & $H \parallel ab$ \\ [-1ex]
    \raisebox{1.5ex}{SC1} & \raisebox{1.5ex}{1.62}& 0.46
    & 0.52 & 252 & $H \parallel c$ \\ [0.5ex]
    \hline
     &  & 0.62 & 0.58 & 238 & $H \parallel ab$ \\ %[-1ex]
    \raisebox{1.5ex}{SC2} & \raisebox{1.5ex}{1.35}& 0.35
    & 0.33 & 316 & $H \parallel c$ \\ [1ex]
   \hline\hline
   \end{tabular}
\end{table}

Our systematic investigations have indicated that BaPt$_2$As$_2$ undergoes a structural phase transition at $T_s\simeq $275 K and two subsequent superconducting transitions at $T_{c1}\simeq1.67$ K and $T_{c2}\simeq 1.33$ K. The structural distortion at $T_s$ is likely driven by a CDW instability. In comparison with the iso-electric compound SrPt$_2$As$_2$~\cite{Fang2012,Kudo2010}, both $T_s$ and $T_c$ are significantly reduced in BaPt$_2$As$_2$. On the other hand, BaPt$_2$As$_2$ and SrPt$_2$Sb$_2$~\cite{Motoharu2013} show similar transition temperatures for both $T_c$ and $T_s$. These observations are in contrast to many other CDW superconductors. For example, in the intensively investigated dichalcogenides~\cite{Gabovich2001}, superconductivity is usually enhanced while the CDW state is suppressed by pressure or doping. Since the electronic states of these 122-type CDW superconductors are expected to be similar, such a distinct behavior might originate from the different electron-phonon couplings and/or spin-orbit couplings arising from the heavier atomic mass of either Ba or Sb. The proportionality of $T_s$ and $T_c$ probably suggests that the CDW state and superconductivity are induced by the same interactions in these compounds. In order to reveal the nature of the structural transition and its interplay with superconductivity in BaPt$_2$As$_2$, further research efforts, e.g., calculations of band structure, pressure study as well as spectroscopic measurements, are highly demanded. Elucidations of the CDW order, superconductivity and their interactions in these compounds may shed new light on other systems, such as the high $T_c$ cuprates.

\section{Conclusion}

In summary, we have successfully synthesized the BaPt$_2$As$_2$ single crystals and studied their structural and physical properties. A structural phase transition, from a CaBe$_2$Ge$_2$-type tetragonal structure (space group P4/nmm) to an orthorhombic phase (space group Pmmn), is observed upon cooling down to $T_s\simeq275$ K. Similar to SrPt$_2$As$_2$~\cite{Imre2007,Fang2012}, such a structural distortion should be associated with a CDW instability. Bulk superconductivity was found in this compound at lower temperatures, with two subsequent superconducting transitions at $T_{c1}\simeq$ 1.67 K and $T_{c2}\simeq1.33$ K. The upper critical fields $\mu_0 H_{c2}(T_c)$ of the two superconducting transitions can be scaled in the same WHH behavior, suggesting for the same pairing state of the two transitions. Analyses of the specific heat data at low temperatures favor nodeless BCS-type superconductivity for BaPt$_2$As$_2$. In comparison with SrPt$_2$As$_2$, the structural transition of BaPt$_2$As$_2$ lies just below room temperature, allowing us to look into the nature of this transition with various advanced spectroscopy techniques. Thus, BaPt$_2$As$_2$ may provide an ideal system to study the interactions of superconductivity and the CDW state.

\begin{acknowledgments}
We are grateful to C. Cao and Han-Oh Lee for their useful discussions. This work was supported by the National Basic Research Program of China (No. 2011CBA00103), the National Nature Science Foundation of China (No.11174245 and No.11474251) and the Fundamental Research Funds for the Central Universities.
\end{acknowledgments}

\end{document}